\documentclass[a4paper,12pt]{article}
\usepackage{algorithm}
\usepackage{algorithmic}
\usepackage{citesort}
\usepackage[numbers]{natbib}
\usepackage{verbatim}
\usepackage[pdftex]{graphicx}

\usepackage{ifthen}
\usepackage{amssymb}
\newboolean{showcomments}
\setboolean{showcomments}{true}
\ifthenelse{\boolean{showcomments}}
  {\newcommand{\mynote}[2]{
    \fbox{\bfseries\sffamily\scriptsize#1}
    {\small$\blacktriangleright$\textsf{\emph{#2}}$\blacktriangleleft$}
   }
  }
  {\newcommand{\mynote}[2]{}
  }

\def \DSpace{{\tt DSpace }}
\def \FindBugs{{\tt FindBugs }}
\def \DSpacens{{\tt DSpace}}

\def \uses{\relation{uses }}

\newtheorem{defn}{Definition}[section]

\newcommand{\dcons}[1]{\texttt{\textbf{#1}}}
\newcommand{\code}[1]{\texttt{#1}}
\newcommand{\relation}[1]{\texttt{#1}}
\usepackage{listings}
\title{The Case for Explicit Coupling Constraints}
\author{Mikal Ziane\footnotemark[1], Mel \'{O} Cinn\'{e}ide\footnotemark[2]}

\begin{document}

\maketitle

%\def\cop{Copyright \copyright\ 2000 John Wiley \&\ Sons, Ltd.}

% Proper dates/numbers go on next line:
%\SMR{1}{7}{00}{00}{2000}

% \runningheads{Mikal Ziane, Mel \'{O} Cinn\'{e}ide}
%{coupling constraints to Facilitate Software Evolution}

%\longaddress{
%- Laboratoire d'Informatique de Paris 6, Paris, France.\\
%- School of Computer Science and Informatics, University College Dublin, Ireland.\\
%}

%\corraddr{LIP6, Paris, France.}

\footnotetext[1]{E-mail: mikal.ziane@lip6.fr}
\footnotetext[2]{E-mail: mel.ocinneide@ucd.ie}

%\received{30 May 2012} \norevised{} \noaccepted{}

\begin{abstract}

A software element defined in one place is typically used in many places. When it is changed, all its occurrences may need to be changed too, which can severely hinder software evolution. A general approach to this problem consists in avoiding to make elements depend on elements which change more rapidly and rather make them depend on more stable abstractions. Encapsulation is a special case of this approach and is supported, in various forms, by most modern programming languages. Unfortunately, as is shown in this paper, this is not enough to express all the constraints that are needed to decouple programming elements that evolve at different paces.

In this paper we show that:
\begin{itemize}
\item A language can be defined to easily express very general coupling constraints.
\item Violations to these constraints can be detected automatically.
\end{itemize}
We then demonstrate several places where the need for coupling constraints arose in open-source Java projects. These constraints were expressed in comments when explicit constraints would have enabled automatic treatment.

% The copyright notice:
%~\cop

\end{abstract}

%\keywords{module decoupling; software evolution; encapsulation; ...}

% ----- INTRODUCTION -----
\section{Introduction}\label{Intro}

% This section is structured as follows:
% context
% problem
% solution

% Here's the context...

The importance of software coupling has been appreciated since the early 1970s when the pioneering work on modular decomposition and structured design was performed \citep{Parnas72, Stevens74}. When modules are loosely coupled, a change in one module is not likely to require that changes be made to other modules. When the reverse is the case, i.e., when a high degree of coupling exists between modules, the result is that maintenance work tends to cause more source code modifications, and indeed an increased error rate \citep{Kemerer95}.

Encapsulation is a particular case of restricting coupling which is supported by modern programming languages but more general cases of coupling restrictions underly a large number of object-oriented design principles and design patterns. Many of the original \emph{Gamma et al} design patterns \citep{Gamma95} can be used to decouple program elements from each other \citep{Ziane03}. An example of a widely-accepted design principle that is fundamentally to do with coupling is the Dependency-Inversion Principle \citep{Martin02}. Its goal is to prevent  high-level modules from depending on low-level modules, so that low level modules can change without causing a ripple of changes up through the higher-level modules.

\subsection*{The problem}

In spite of the recognized importance of restricting software coupling, it is imperfectly supported in current programming languages.

Consider for example the Factory Method and Prototype design patterns \citep{Gamma95} which both aim at shielding client code from changes in concrete classes (the \emph{concrete products}) that they however need to instantiate.

The solution of both patterns thus includes indirect means to instantiate the product classes: the \texttt{clone} method in the Prototype pattern, and a so-called \emph{factory method} in a class hierarchy parallel to that of the products (the \emph{creators}) in the Factory Method pattern.

These patterns however do not include means to prevent direct instantiations of the concrete products to prevent correct implementations of the patterns to become corrupted. It is thus natural to try and use one's favourite programming language to do that. Alas, in many cases this does not work.

In Java, for instance, the concrete product classes may be put in a different package from their abstract class and given the default (package) rather than public visibility. This is acceptable in simple occurrences of the Prototype pattern where no client class or method is given special privilege to instantiate some or all of the product concrete classes. But with the Factory Method design pattern this would force to put the so-called \emph{concrete creator} classes in the same package as the concrete products.

More generally, relying on packages to enforce coupling restrictions is not a general solution. For example, if some class C needs privileged access to a non-public class A of a package it must be put in the same package as A. But access to C itself may have to be restricted too, so that C must be given non public visibility and its direct clients, including say CC, must be put in the same package as C. Since packages cannot intersect this means putting CC in the same package as A, while one may not want CC to be granted access to A.

What is needed is a general solution to prevent an arbitrary set of program elements from using other program elements (the \emph{services}) while programming languages only offer very restrictive means of expression. With the \texttt{friend} keyword of C++, for instance, the (revealed) services are all the members of the class bearing the friend clause, not just a selection of them. Moreover, the elements that are allowed to access the services must be explicitly named one by one which makes the friend clause itself very fragile. Finally, this explicit naming of the elements which are granted access to the class members makes the class statically depend on them! \footnote{Suppose for instance that all the subtypes of some type, including some that are not yet defined, need to be friends. Removing a subtype break the friend clause and the class definition! Adding a subtype makes it incomplete with respect to the original intention.}.

% First, neither the program elements which are to be used nor the program elements which are allowed to use them can be defined by a for all the members of the class C bearing the friend declaration can be accessed by f when one may just want to let f use some but not all of the members of C. \item Second, the friend declaration makes class C statically depend on f! If f is removed then class C must be changed.\item Third, the friend declaration itself is fragile \end{itemize}

%% MAYBE DEFER THIS DISCUSSION TO ANOTHER SECTION AND POINT TO IT IN THE INTRODUCTION

In the absence of language support for this type of coupling constraint, two other options can be used. The original programmer or system architect who intends two modules to be decoupled, and to remain decoupled, can express this either in documentation, or by relying on the insight of future maintenance programmers to understand the intention of their design. Neither solution is ideal. Comments are often ignored\footnote{Indeed, Agile practices suggest that comments can be a sign of poor design, and that where possible the design should be refactored to make the comment unnecessary \citep{Fowler99}. In the case of the comments under discussion, the goal is to alert maintenance programmers to avoid particular couplings; the design may be completely adequate and no refactoring required.}, and maintenance programmers cannot be relied upon to appreciate  and observe the coupling limitations implied in the original design.

The fact that comments are used in software projects to warn against some couplings is further evidence that current language support is incomplete. But the fact that design decay still occurs over time suggests that tool support is needed to prevent it which is not possible with implicit or informal coupling constraints.

%\begin{figure}
%\begin{center}
%%\includegraphics{FactoryMethod.eps} % This .eps has a bug in it :-(
%\includegraphics[width=.8\linewidth]{FactoryMethod_PNG} %taken from wikipedia
%\caption[FactoryMethod]{Structure of the Factory Method Pattern }\label{FactoryMethod}
%\end{center}
%\end{figure}

%\mel{wrong labels \& relationships, needs fixing!}

\subsection*{The solution}

In order to address these issues, we introduce the concept of an explicit \emph{coupling constraint}. Here "explicit" not only means that the constraint is fully and clearly expressed but also that it can be checked automatically. In the rest of this document we shall assume, if not otherwise stated, that coupling constraints are explicit.

A coupling constraint expresses the requirement that some program elements (package, class,  method ...) should not statically depend on other program elements, typically when the former elements are expected to vary more frequently than the latter ones.
Coupling constraints are defined by the original programmers or more likely by the system architect and may be automatically checked whenever the software is later updated by a maintenance programmer. If a coupling constraint is violated, the maintenance programmer will need to refactor the code or to relax the constraint.

In a simple scenario of the already mentioned Factory Method and Prototype patterns, a simple (and very strict) constraint could first be informally stated as thus: hide the concrete product classes. This would be similar to giving these top-level classes the private visibility which is however not possible in programming languages such as Java, C++ or C\#.

This simple constraint is compatible with a solution to the Prototype pattern if all the instantiations of a concrete product class (including the creation of the prototypes) occur in the scope of this very class, which can be a bit tricky to achieve. Otherwise, the constraint must be relaxed to let the prototypes be instantiated. In the case of the Factory Method pattern the constraint must be relaxed to let each concrete creator class use the corresponding concrete product class.

 %all the implementations of the {\code Product} interface should be hidden from the {\code Creator} class. This includes the existing  {\code ConcreteProduct} class, and all future implementations that may be added. Less obviously, the {\code Product} interface should also be decoupled from the {\code ConcreteCreator} classes. This permits the {\tt ConcreteCreator} classes to create instances of the implementing classes of {\code Product}, but not to invoke their methods. The details of the coupling constraints required to achieve this will be provided in subsection \ref{DC}.

The remainder of this paper is structured as follows....

In section \ref{SDAG} we describe our notion of coupling constraint in detail, present the graphical technique we use to depict coupling constraints, and present a precise definition of the coupling constraints used in this paper. In section \ref{Eval} we evaluate our work by seeking examples of coupling constraints in open source software and demonstrate how these can be detected using our tool, Lutin. In section \ref{RW} we review related work in the area of software coupling. Finally, in section \ref{CnFW} we present our overall conclusions and discuss future work in this area.

\section{Static dependencies and access graphs}\label{SDAG}

What is a static dependency to an entity? We assumed that, aside from the mere duplication of code which we are not addressing in this paper, a static dependency involves using an entity e by its name. If e is removed or even changed, each occurrence of its name may lead to compilation errors\footnote{We dot not currently take into account occurrences in literal strings nor in comments.}.

We are thus only considering entities with a name which, following Java's terminology \citep{JLS}, we call \textbf{declared entities} (packages, classes, interfaces, class members ...).
Names may be partially implicit in programs but we assume that a deterministic procedure can statically (i.e. before execution) produce a fully qualified name from a partial name and its context.

Static names may still be ambiguous with respect to inheritance polymorphism, which is resolved by dynamic binding, but this is intentional as our goal is to pinpoint \textbf{static} dependencies. We thus introduce the following definitions.
% If the subtypes of O change (e.g. subtypes are added or removed) but O itself is unchanged, the method call o.m() should still be valid, at least if it is not, this is not our concern here. So, the dynamic type of o does not interest us. In fact, to decouple two pieces of code it is common (e.g. in several GoF patterns) to replace static dependencies by dynamic binding.

\begin{defn}[Owner and declaration scopes of an entity]\ \\
Each declared entity is owned by a scope which, intuitively, is the smallest scope that strictly includes the declaration of the entity\footnote{The Java Language specification defines the scope of an entity as "the region of the program within which the entity [...] can be referred to using a simple name, provided that it is visible"\citep{JLS}\S6.3.}. The declaration of the entity is also typically a scope itself: the declaration scope of the entity.\footnote{The term "definition scope" would be better suited for languages like C and C++ where a declaration is not the same as a definition in which case declared entities should probably be renamed as defined entities.}
\end{defn}
For instance, the declaration scope of a method is the whole method declaration including its body, if there is one, while the owner scope is the class or the interface bearing the declaration.

\begin{defn}[Static dependency to a declared entity]\ \\
A static dependency to a declared entity \texttt{e} in a program \texttt{P} is any occurrence of the name of \texttt{e} in \texttt{P}.
%outside of the declaration scope of e
An entity \texttt{c} statically depends on \texttt{e} when there is at least one static dependency to \texttt{e} in the declaration scope of \texttt{c}.\footnote{Trivial dependencies such as the mandatory occurrence of a name in its own declaration are omitted.}
\end{defn}

%Note that according to these definitions an entity depends on itself and on its sub-entities.

%The reason for not taking into account the occurrences of the name of an entity $e$ inside the declaration scope of $e$ itself, is that this can hardly be considered (potentially harmful) coupling.

% Discussion: some dependencies can be inferred. For instance if Car and Bike are mentioned then when the type Motocycle is introduced it likely must be taken into account.
% Discussion: in addition to deleting e, changing the type of e may break the occurrences of n.
% Discussion: clones of the same code fragment are akin to dependencies albeit with no explicit name for e: e is used by duplication not by its name.

\subsection*{Access graphs}
In order to define coupling constraints as independently as possible from any particular programming language, programs are abstracted by a relational structure: an access-graph.
Access graphs also make it easier to reason about static dependencies in programs, by focusing on the relevant concepts.

\textbf{Nodes} in access graphs denote declared entities while \textbf{relations} either bind entities which use other entities or are useful to qualify which entities are allowed to use other entities. Several dependencies to the same target entity that occur in the same source entity will appear as a single edge from the node of the source entity to that of the target entity.

%For instance, multiple dependencies to the same target in the body of a method will be grouped as a single dependency from the node which represents the method.
% The individual occurrences in a program can still be recovered, say to pinpoint targets for refactoring.

The central relation of access graphs is the \textbf{\uses} relation.
\begin{defn}[The \texttt{uses} relation of a program]\ \\
Let \texttt{P} be program.
A declared entity \texttt{c} of \texttt{P} \texttt{uses} another declared
entity \texttt{e} of \texttt{P} when \texttt{c} statically depends on \texttt{e}.
\end{defn}

% The nodes of an access graph denote declared entities and the edges denote relations typically including inheritance, inclusion (which binds a scope to its elements) and the abstract access relation uses. This abstract access relation must be an abstraction of the

\begin{defn}[Access Graph of a program]\ \\
An access graph \\*
$g=<Nodes,Relations,uses_g>$ of a program \texttt{P} is a graph whose nodes are declared entities of \texttt{P}
and with a special relation $uses_g$ which is the \texttt{uses} relation of \texttt{P} restricted to these entities.
\end{defn}

Coupling constraints will be defined below as logical formulas that forbid some \uses edges in access graphs. In addition to the \uses relation, other relations (e.g. inheritance or aggregation) are typically included into access graphs to qualify what \uses edges are allowed or forbidden: the only requirement is that these relations can be automatically computed from a given program.

Access graphs are useful to define the semantics of coupling constraints, to reason on them, and to display what depends on what or which dependencies violate a given coupling constraint.
Note however that access graphs may be displayed partially to improve readability.
%In particular some \relation{uses} edges may not be displayed if they can be deduced from other relations.
%For instance the dependencies of an entity to itself or its sub-entities are not displayed in this paper.

\lstset{basicstyle=\small}
\begin{figure}
\centering
\begin{lstlisting}[name=ImageMgr,
    numbers=left,
    numberblanklines=false,
    frame=trl,
    belowskip=0pt
    ]
// coupling constraint: hideScope('ImageDoc').

public class ImageDoc {
  public ImageDoc() {name="my Image";}
  public String getName() {
    return name;
	}
  private String name;
}
\end{lstlisting}
\begin{lstlisting}[name=ImageMgr,
    numbers=left,
    numberblanklines=false,
    morekeywords={ImageDoc, getName},
    frame=rlb,
    aboveskip=0pt,
    ]
public class ImageMgr {
  private ArrayList<ImageDoc> images;
  public ImageMgr() {images = new ArrayList<ImageDoc>();}

  public void display () {
    for (ImageDoc d : images)
      System.out.println(d.getName());
  }

  public void addImage() {
    images.add(new ImageDoc());
  }

  public static void main(String[] args) {
    ImageMgr mgr = new ImageMgr();
    mgr.addImage();
    mgr.display();
  }
}
\end{lstlisting}
\caption{Image Manager example} \label{fig-imageMgr}
\end{figure}
\normalsize

Consider the Java program of figure \ref{fig-imageMgr}. The $ImageMgr$ class manages Image documents (instances of the $ImageDoc$ class). In order to prepare the evolution of the program to support different kinds of documents, the $ImageMgr$ class should not depend on the $ImageDoc$ class but on a more stable abstraction.

\begin{figure}
\begin{center}
%\scalebox{0.3}{\includegraphics*{vim.eps}}

\includegraphics[height=.55\textheight]{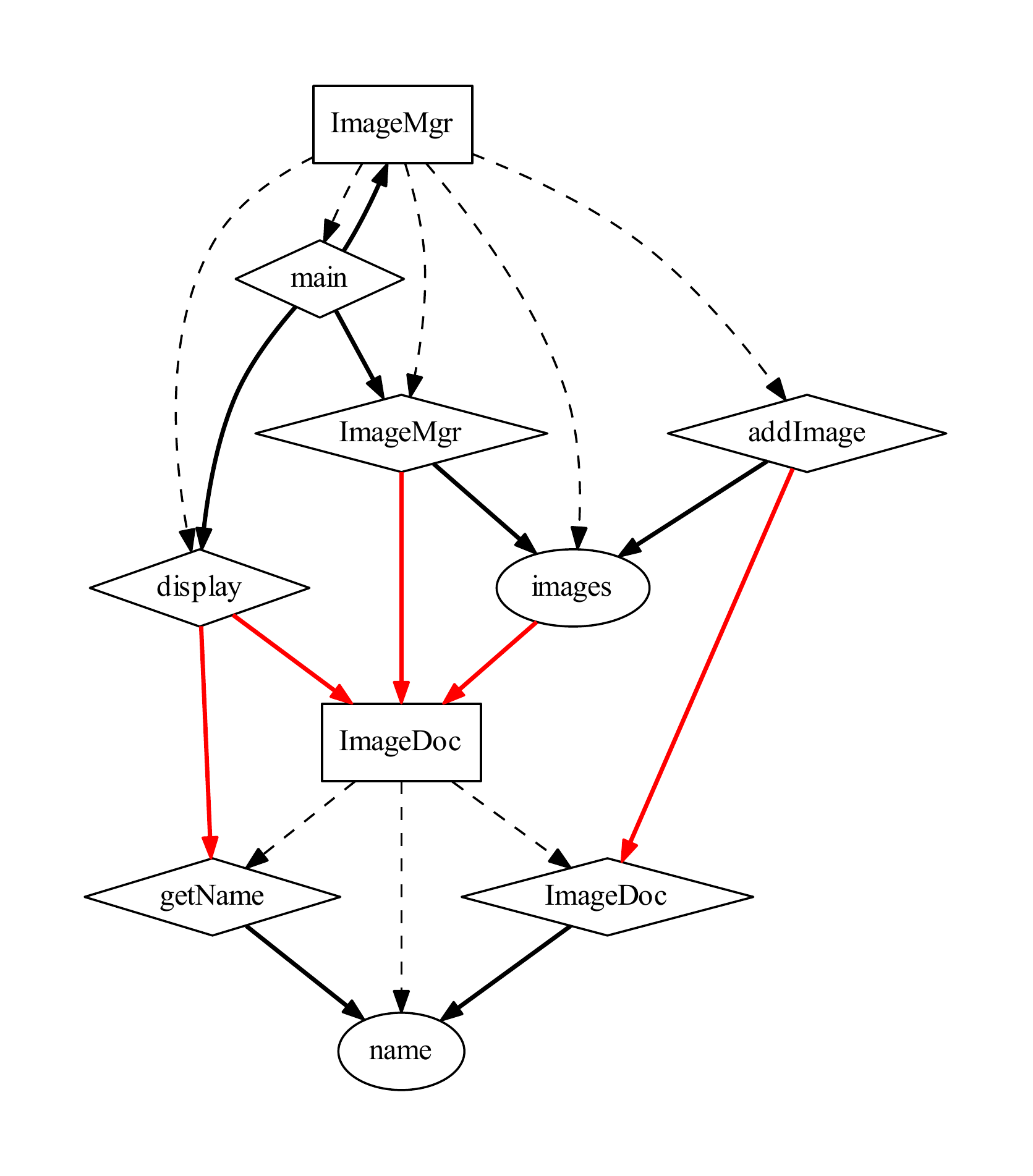}
\caption{Access graph of the Image Manager program.}
\label{ViolGraph}
\end{center}
\end{figure}

The access graph of figure \ref{ViolGraph} was computed by a tool, Lutin, described in section \ref{implem}. It displays the \uses relation as \emph{full lines} and the \texttt{contains} relation (see section \ref{fol}) as \emph{dashed lines}. \emph{Squares} are classes or packages, \emph{diamonds} are methods or constructors and \emph{ovals} are data members. The \textbf{red edges} are dependencies that violate a coupling constraint as will be explained below.

Six dependencies are pinpointed as problematic with respect to the (yet informal) coupling constraint that the $ImageMgr$ class should not be used (directly) from the class $ImageDoc$. The name of the class $ImageDoc$ is used (as a type name) in the $ImageMgr$ constructor, in the $ImageMgr.display$ method as well as in the declaration of the $ImageMgr.images$ attribute. In addition, $ImageMgr.images$ method uses the $ImageDoc.getName$ method and the $ImageDoc.addImage$ method uses the constructor $ImageDoc$ but also while doing that, also the name of the class $ImageDoc$ itself.

\section{Coupling constraints}\label{CC}
Coupling constraints are now defined as logical expressions that forbid some \relation{uses} edges in access graphs. The only couplings that will be considered are those that are compatible with the syntax and especially the access rules of a programming or a modeling language.
Thus, coupling constraints further restrict the couplings among the syntactically correct ones.

It is beyond the scope of this paper to discuss which changes in a given access graph, if any, are able to fix the violations coupling constraints. However, the point of coupling constraints is to guide refactoring by rigourously determining which refactoring combinations are intended among the possible ones.

Coupling constraints can be interpreted at two different levels of abstraction: access graphs or programs. For instance,  $hiddenFrom(b,a)$ first means that in the considered access graph an  \relation{uses} edge from $a$ to $b$ would be \textbf{incorrect}\footnote{Remember though that the forbidden \uses edges are called \emph{incorrect} only with respect to a given set of coupling constraints and this has nothing to do with behavior preservation or the syntax of the programming language which are always assumed to be respected.}. An incorrect edge is a potential target to apply a refactoring transformation. In this paper, such edges are displayed in red. Second, given a program P, $hiddenFrom(b,a)$ means that in P the occurrences of the name of $b$ in the scope of $a$ are incorrect.

\subsubsection*{The need for a logical language}
An elementary formula like $hiddenFrom(b,a)$, is typically not enough to express a useful constraint for at least two reasons.
\begin{itemize}
\item First, a given program element typically needs to be hidden from a large number of other elements, possibly including elements which will be added to the program after the constraint was defined.
\item Second, $hiddenFrom(b,a)$ forbids $a$ from using $b$ but says nothing about the nested elements in $a$ or $b$. Access to sub-elements often need to be restricted when access to their owners is.
\end{itemize}
We thus now introduce a first-order logical language to express coupling constraints and then higher-level predicates to ease the declaration of the most common constraints.
%This is seldom a problem for $a$ because if there is a static dependency to an entity of $a$ the name of $a$ has typically been mentioned too. For instance if $o.m()$ statically denotes a method of class
%Moreover, in access graphs, edges are attached to the nodes of the smallest scope in which the forbidden names occur. If A is a class, the \relation{uses} edges will most often start from the nodes of the methods of A rather than from the node of A itself.

\subsection{First order language}\label{fol}
The first-order language that we propose to define coupling constraints includes:
\begin{itemize}
\item a set of constants,
\item a set of variables,
\item the usual logical symbols,
\item a signature: a set of binary relational symbols including a special relation \uses and a set of predicates including three special binary predicate \relation{hiddenFrom}, \relation{hideFrom} and \relation{canSee}.
\end{itemize}

Given an access graph, this language can be interpreted this way: the variables and constants denote nodes or sets of nodes of the graph and binary relations denote sets of edges. The \relation{uses} relation of the language denotes the uses relation of the graph. Predicates are interpreted the usual way.

The point of the language is to express which \uses edges are correct and which are not. This is done by the introduction of constraints from which \relation{hiddenFrom} facts can be deduced.

\begin{defn}[violation of a set of coupling constraints]\ \\
Given a set of coupling constraints $C$, a $uses(a,b)$ edge of an access graph is a violation of $C$ if $C \Rightarrow hiddenFrom(b,a)$.
\end{defn}

\subsubsection{Dealing with exceptions}

In order to allow for the introduction of local exceptions to global decoupling policies that should remain unchanged, it is advised to use the softer \textbf{\relation{hideFrom}} predicate rather than \relation{hiddenFrom} directly. Exceptions can then be introduced using the \textbf{\relation{canSee}} predicate. Note that all the high-level predicates defined below are defined using \relation{hideFrom} rather than \relation{hiddenFrom}.

The following axiom defines the relationship between the three predicates.

\begin{defn}[hiddenFrom axiom]\ \\
$hideFrom(b,a) \, \wedge \neg canSee(a,b) \Rightarrow hiddenFrom(b,a)$
\end{defn}

Using \relation{canSee} should be done very sparingly, though, as it bypasses all the constraints that rely on \relation{hideFrom}. A more cautious way to introduce exceptions consists in including them directly in coupling constraints as allowed by most of the high-level predicates defined in this document. To avoid making constraints depend on specific nodes, one can define them using variables that denote sets of nodes.

\subsubsection{Other low-level relations and predicates}
The simplest way to forbid node $a$ to use node $b$ is simply to declare the $hideFrom(b,a)$ constraint. First order constraints can also be written the usual way to hide an arbitrary set of node from other nodes. For this purpose, functional or relational symbols can be added to the logical language as long as they are unambiguously defined on access graphs and programs.
For instance, the \relation{isClass} unary predicate can be added to denote nodes that are classes.

Two relations are particularly useful: \relation{contains} and \relation{isA}. \relation{\textbf{Contains}} can be given a rather generic definition and is quite convenient to hide a whole scope including the elements defined in it. In practice though, it is often \relation{$\mathbf{contains^*}$}, the reflexo-transitive closure of \relation{contains}, that is actually used in coupling constraints.

The precise definition of \relation{\textbf{isA}} depends on the programming language but it is quite important for two reasons. Firstly, it is convenient to forbid the use of all the subtypes of a given type including those that have not been defined yet.

Secondly, and more importantly, the \relation{isA} relation is central to solving coupling problems trough dynamic binding. If a method call $a.m(...)$ is forbidden, where $a$ is of static type $A$, then a common refactoring consists in declaring $a$ to be of type $T$, where $T$ is a super type of $A$ which either exists or needs to be inferred (with the appropriate methods) and introduced. So, while the \relation{isA} relation is not absolutely necessary to define coupling constraints it is often essential to their satisfiability.

\begin{defn}[contains]\ \\
A declared entity $e$ \relation{contains} a declared entity $e'$ iff $e$ is the owner scope of $e'$.
\end{defn}

\begin{defn}[isA]\ \\
A declared entity $s$ isA $t$ iff both are types and $s$ is defined as a subtype of $t$. This implies that wherever an expression of type $t$ is expected, an expression of type $s$ may occur.
\end{defn}

\subsection{Higher-level predicates and relations}

The \relation{hideFrom} predicate is quite low-level and it is often more convenient to rely on higher-level predicates and relations. The following definitions are given in first-order logic and have been implemented in prolog (see the Appendix).

\begin{defn}[Virtual scopes and virtual\_contains]\ \\
A virtual scope is an arbitrary collection of declared entities that are put together so that they can easily be considered as a whole in coupling constraints. The virtual scope becomes a node that virtually contains its elements.

\small{
$virtualScope(s,elements) \equiv \:
\\
\exists node \, node = s \wedge \forall e \,  e \in elements \to \: virtual\_contains(s,e)$
}
\end{defn}
An example of virtual scope is given in section \ref{DC_Package} where one of the layers of a layered architecture is not a scope but a collection of scopes.

In order to deal with virtual scopes and actual scopes uniformly in constraints it is convenient to introduce a generalize contains relation which also supports set (or any kind of collection) membership so that sets of entities and single entities can be dealt with uniformly too.

\begin{defn}[generalized contains]\ \\
\small{
$gContains(a,b) \equiv (b \in a) \lor contains(a,b) \lor virtual\_contains(a,b)$
}
\end{defn}

A constraint $hideScope(s, facades, interlopers, friends)$ hides a scope s, except for a set of facades, from a set of scopes (the interlopers) except from a set of friends which are not interlopers after all. Simpler versions of this predicate are also convenient:
\begin{itemize}
\item hideScope(s) that hides a scope s from anything outside of it (i.e. from anything that s does not gContains),
\item hideScopeBut(s, facades) that hides s except for a set of facades,
\item hideScopeFrom(s, interlopers) that hides s from a set of scopes (the interlopers),
\item hideScopeButFrom(s, friends) that hides s but from a set of scopes (the friends).
\end{itemize}

\begin{defn}[hideScope]\ \\
\small{
$hideScope(scope,facades,interlopers,friends) \equiv
\\
\forall e \, \forall i \,
    ( gContains^*(s,e) \land \,
      gContains^*(interlopers,i) \land \\
      \neg gContains^*(facades, e) \land
      \neg gContains^*(friends, i) \land
      \neg gContains^*(s, i)
     ) \\
     \to hideFrom(e,i)$
}
\end{defn}

%\begin{defn}[hideSubTypes]\ \\
%The subtypes of a type are hidden from all (except from themselves).
%
%\small{
%$HideSubTypes(T)  \equiv \forall S \: isA^+(S,T) \to HideScope(S)$
%}
%\end{defn}

\subsection{Using coupling constraints and access graphs}
Consider again the program of \ref{fig-imageMgr}.
Declaring an explicit coupling constraint works in two ways. First it makes explicit in an unambiguous way the decoupling intention of the architect of the application. Second it allows the automatic detection of the dependencies that do not comply with this constraint.

Depending on the intention of the developer the \code{ImageDoc} class could be hidden either from the \code{ImageMgr} class specifically or from every name space (but itself) in the program. Both constraints are equivalent for the program we are considering but if more classes are added it will be necessary to clarify which ones can access \code{ImageDoc}. Let us assume that the second option has been chosen and that the following constraint is added: \textbf{\dcons{hideScope(ImageDoc)}}

This not only means that the \code{ImageDoc} identifier cannot be used outside its own scope, but that the identifiers defined in the \code{ImageDoc} scope cannot be used outside \code{ImageDoc} either. For instance, the occurrence of \code{getName} line 16 is not allowed because as the static type of the \code{d} variable is $ImageDoc$, it statically denotes the \code{ImageDoc.getName} method. The bold identifiers in figure \ref{fig-imageMgr} are those whose occurrence is not allowed by the coupling constraint.

On the  access graph of figure \ref{ViolGraph} each red edge denotes at least one violation of the coupling constraint.

\subsection{Implementation}\label{implem}
Lutin is a three-step tool which takes as input Java software (a jar archive) and a coupling constraint written in Prolog. It produces several versions of the access graphs of the software, in different formats. 

The front-end of Lutin statically analyses \footnote{Different versions of Lutin have relied on different parsing and analysis toolboxes such as StrategoXT \cite{Stratego08}. The StrategoXT version is available at http://pagesperso-systeme.lip6.fr/Mikal.Ziane/lutin/stable/site/.} the Java code and generates a Prolog representation of the access graph.

The second step is an independent Prolog program which checks a coupling constraint on an access graph and produces a modified access-graph in which the violations are highlighted. The coupling-constraint language is thus implemented here as an embedded DSL where Prolog is the host language.

The last step displays (currently relying on GraphViz) displays the modified access graph.

%------------------------------------------------------------------------------------
\section{Evaluation}\label{Eval}
%------------------------------------------------------------------------------------

Our approach to evaluation is to demonstrate firstly that there is a need for coupling constraints, then to show how these coupling constraints can be detected, and finally to evaluate our approach to detection on an open source example.

To determine if there is a need for coupling constraints, we consider what a programmer might do if they encounter the need for a coupling constraint in their code that cannot be expressed in the programming language itself. We hypothesise that a diligent programmer might express it as a comment to alert future maintenance programmers not to create the coupling in question. Such comments, if discovered, could provide insight into what type of coupling constraints are required in practice.

For our case study, we examined in detail one medium-sized open-source Java application, namely \DSpace version 1.5.1 \citep{DSpace08}. \DSpace is an open-source Content Management System written primarily in Java. It was originally developed jointly by MIT Libraries and Hewlett-Packard before being released into open source. It comprises just under 100 KLOC of Java code and contains 75 KLOC of comments, and so provides a rich domain in which to seek comments that relate to coupling constraints.

Our aim was to find comments that express the need for coupling constraints. We filtered the comments initially using coupling-related terms namely ``access," ``coupling," ``coupled," ``depend," ``know," and ``visibility." We then inspected each comment manually to determine if it was in fact related to coupling or not. The results of this analysis are presented in section \ref{DC_DSpace}. In subsection \ref{DetectingViolations} we illustrate how we can detect violations of these coupling constraints and finally, in subsection \ref{Discussion}, we discuss our results.

\subsection{Coupling constraints found in \DSpace} \label{DC_DSpace}

In the following subsections we present examples of the type of coupling constraints that were found in \DSpace and, in each case, show how the constraint can be represented in our formal notation. All the evidence presented here is based on comments found in the source code, except for the first example in section \ref{DC_Package}, which is based on \DSpace design documentation.

\subsubsection{Decoupling from a Package} \label{DC_Package}

Decoupling between packages is of the upmost importance as it relates to the system architecture, and problems at this level cannot be easily resolved with local measures.
\begin{figure}[ht]
\begin{center}
\includegraphics[width=\linewidth]{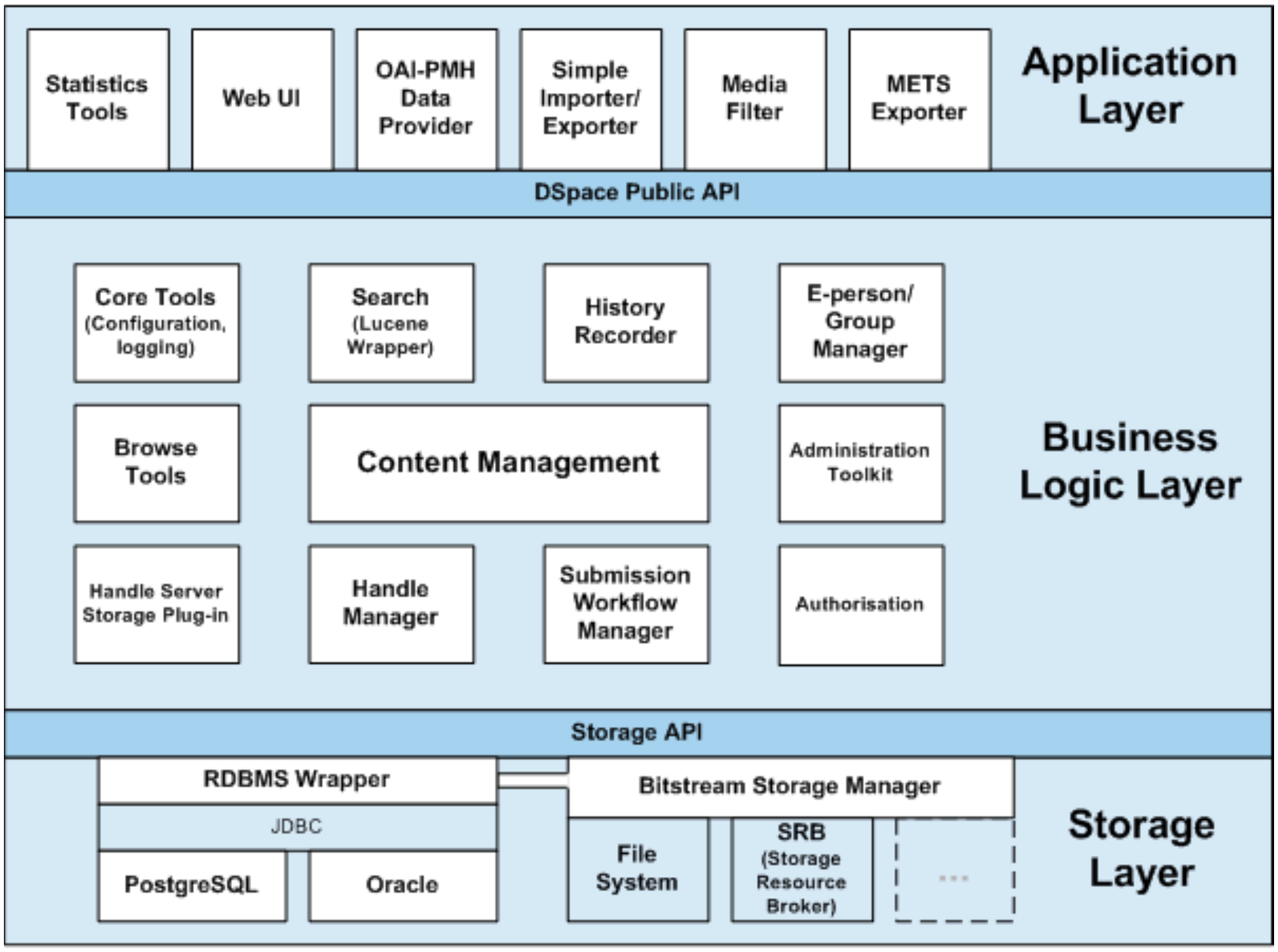}
\caption{Logical Architecture of the \DSpace Application}
\label{fig:DSpaceArchitecture}
\end{center}
\end{figure}
As can be seen in Figure \ref{fig:DSpaceArchitecture}, \DSpace uses the standard 3-tier layered architecture. A key aspect of this architecture is that each package (layer) should use only the package immediately below it. This implies that a package should be decoupled from all the other packages, except the package immediately below it. These coupling constraints can be expressed thus:

\begin{small}
\begin{verbatim}
virtualScope('org.dspace.business',
	['org.dspace.administer',
	'org.dspace.authenticate',
	...]).
hideScopeFrom('org.dspace.app',
              ['org.dspace.business','org.dspace.storage']).
hideScopeButFrom('org.dspace.business',
                 ['org.dspace.app']).
hideScopeButFrom('org.dspace.storage',
                 ['org.dspace.business']).
\end{verbatim}
\end{small}

The first declaration defines the Business Logic Layer as a virtual scope as it is in fact not a single  package in \DSpace but a collection of packages. The first constraint says that the Business Logic layer and the Storage layer may not use the Application layer. The second constraint hides the Business Logic layer to anything outside its boundaries but the Application Layer. The last constraint does similarly with the Storage layer which can only be accessed from outside its boundaries by the Business Logic layer.

Since the layered architecture is common, we also introduced a higher-level predicate, \emph{layers}, so that the five constraints above could be replaced by just one:

\begin{small}
\begin{verbatim}
layers(['org.dspace.app',
    'org.dspace.business',
    'org.dspace.storage']).
\end{verbatim}
\end{small}

Another example of decoupling from a package was discovered in the {\tt METSExport} class, where the following comment appears:
\begin{quote}
\emph{We don't pass up a MetsException, so callers don't need to
know the details of the METS toolkit.}
\end{quote}
%The METS (Metadata Encoding and Transmission Standard) schema is a standard XML language for encoding object metadata within a digital library.
The {\tt METSExport} class provides high-level wrapper methods to access the METS toolkit, and the comment expresses the constraint that classes that use the METS toolkit should not be exposed to any exceptions defined by the toolkit. More generally, it means that the classes in the METS package, except for {\tt METSExport}, should be hidden from the other classes of the Application layer. This can be expressed as follows:
%More generally, it states that the application classes should be decoupled from the classes in the METS package, accessing them only using the {\tt METSExport} class. This implies a constraint that decouples the \DSpace classes, other than {\tt METSExport}, from the set of METS classes, which can be expressed using the {\tt SoleCoupling} constraint as follows:
\begin{small}
\begin{verbatim}
hideScopeBut('org.dspace.app.mets', ['METSExport']).
\end{verbatim}
\end{small}

%Note that a maintenance programmer who updates one of the \DSpace classes to use the METS library would have to import explicitly the METS package, and this might suggest to them that the coupling is not appropriate.
%This explicit coupling constraint prevents an inexperienced maintenance programmer from creating this undesirable coupling.

\subsubsection{Decoupling from a Class} \label{DC_Class}

In chapter 6 of the \DSpace documentation, the following comment appears:
\begin{quote}
\emph{The BitstreamStorageManager provides low-level access to bitstreams
stored in the system. In general, it should not be used directly;
instead, use the Bitstream object.}
%... in the content management API since that encapsulated authorization and other metadata to do with a bitstream that are not maintained by the BitstreamStorageManager.}
\end{quote}
This warns programmers not to use the {\tt BitstreamStorageManager} class directly but to use instead the  {\tt Bitstream} class. Looking at this in terms of coupling constraints, what is required is that all \DSpace classes other than the {\tt Bitstream} class should be decoupled from the  {\tt BitstreamStorageManager} class. This can be achieved thus:
\begin{small}
\begin{verbatim}
hideScopeButFrom('org.dspace.storage.bitstore.BitstreamStorageManager',
     ['org.dspace.content.Bitstream']).
\end{verbatim}
\end{small}
If a programmer accidentally uses {\tt BitstreamStorageManager} from another class in the application, a coupling constraint violation will be raised.

Another example appears in the same class, {\tt BitstreamStorageManager}, where the following comment appears:
\begin{quote}
\emph{The dependency on the checker package isn't ideal...}
\end{quote}
On closer inspection, the dependency in question is actually on the class {\tt BitstreamInfoDAO}. In terms of the coupling constraint required here, it is simply a matter that the{\tt BitstreamInfoDAO} class should be hidden from the class {\tt BitstreamStorageManager} which can be expressed thus:
\begin{small}
\begin{verbatim}
hideFrom('BitstreamStorageManager', 'BitstreamInfoDAO').
\end{verbatim}
\end{small}
%We return to this particular example in the context of soft coupling constraints in section \ref{DC_Soft}.

\subsubsection{Decoupling one Method from another} \label{DC_Method}

In the {\tt Bitstream} class, the following comment appears in the {\tt create} method:
\begin{quote}
\emph{...This method ... does not check authorisation; other methods such as {\tt Bundle.createBitstream()} will check authorisation.}
\end{quote}
This implies that the {\tt Bitstream::create} method should not access the method that checks authorisation, namely {\tt authorizeAction} in the {\tt AuthorizeManager} class, because other methods are responsible for performing this check. This decoupling can be expressed thus:
\begin{small}
\begin{verbatim}
hideScopeFrom('AuthorizeManager.authorizeAction', 'Bitstream.create').
\end{verbatim}
\end{small}
This constraint prevents the maintenance programmer from erroneously invoking  {\tt authorizeAction} in the {\tt Bitstream::create} method, believing authorisation to be part of creating a bitstream. If this dependency is created, the subsequent coupling constraint violation  will direct the programmer to seek another solution.

Another example of decoupling from a method is found in in the following comment that appears in the {\tt DAVEPersonEPerson} class:
\begin{quote}
\emph{Give read-only access to the contents of an EPerson object...}
\end{quote}
The purpose of the {\tt DAVEPersonEPerson} class is to serve as an Adaptor \citep{Gamma95} for the {\tt EPerson} class, i.e., to prohibit access to the mutator methods in {\tt EPerson}.
%In Java, this can be achieved by creating an interface that only allows access to the accessor methods. C++ provides better support, by allowing objects to be declared as {\tt const}, thereby permitting access only to operations declared as {\tt const}. This C++ technique relies on the programmer of the class following the discipline of using {\tt const} rigourously, which is frequently not the case.
A coupling constraint can be used to good effect here, in order to state that particular clients of the {\tt EPerson} class are to be decoupled from its mutator methods. This avoids the necessity of creating a new interface, or relying on programmer discipline to preserve the decoupling.

Closer examination of the \DSpace code reveals that the {\tt DAVEPersonEPerson} class has two clients, namely {\tt Item} and {\tt WorkflowItem}. Also, the {\tt Eperson} class contains ten mutator methods, which we refer to as {\tt EPerson\_mutators}. The required coupling constraint can then be expressed:
\begin{small}
\begin{verbatim}
declareSet('EPerson_mutators',['EPerson.setEmail', ...]).
hideSet('EPerson_mutators').
\end{verbatim}
\end{small}
%ForbidCoupling(Item, EPerson_mutators)
%ForbidCoupling(WorkflowItem, EPerson_mutators)
By defining these coupling constraints, we ensure that the client classes are not erroneously updated to access mutator methods in the {\tt Eperson} class. Furthermore, the rather artificial {\tt DAVEPersonEPerson} class can now be deleted from the program as its role has been assumed by these coupling constraints.
%\mel{There is a maintenance issue here that arises when new mutators are added. We can mention here, or address elsewhere.}

\begin{comment}
For the record, the full mutator list for EPerson is:
{\tt setPassword}, {\tt setLanguage}, {\tt setEmail}, {\tt setNetid}, {\tt setFirstName}, {\tt se
tLastName}, {\tt setCanLogIn}, {\tt setRequireCertificate}, {\tt setSelfRegistered} and {\tt setMetadata}.
\end{comment}

\subsection{Detecting violations of \DSpace coupling constraints}\label{DetectingViolations}

In the preceding section we presented evidence from the \DSpace documentation that programmers see the need for coupling constraints and sometimes express them as comments. Due to the lack of language or tool support for coupling constraints, this is the only option open to them. It may be claimed that expressing coupling constraints as comments is an adequate solution. Maintenance programmers will read the comments, take heed of their advice and avoid the undesirable couplings.

To test if this is the case have used two different tools to check the constraints described in section \ref{DC_DSpace}. Both tools have been run on  \DSpace code to detect if the coupling constraint has been observed or not. The main reason for using two tools was that the Lutin prototype was not, until recently, mature enough to deal with software as large as DSpace.

So a first series of experiments were conducted using \FindBugs \citep{FindBugs08}, an open source static analysis tool for Java. More recently, the same series of experiment was then run using Lutin with the same results.

With \FindBugs a specific detector has to be implemented for each constraint.
Such a detector examines a Java program looking for a specific set of patterns or rules by matching program bytecode against a list of specified "bug" patterns. A bug in this context is really a code smell, i.e., an undesirable design construct. The input to each detector is an XML file that provides the necessary parameters.

On the one hand the learning curve of \FindBugs is by far not as steep as that of the analysis tools underlying Lutin. On the other hand, once the analysis frontend of Lutin had been adapted to handle large programs, it was extremely easy to experiment with new constraints in prolog with the fullpower of the constraint language.

%\FindBugs uses a large pre-defined library of bug detectors, but also permits the user to define their own detectors.

%\FindBugs uses the Visitor pattern to perform its analysis. In order to define a new detector, the {\tt BytecodeScanningDetector} Visitor is subclassed and the appropriate methods overridden to perform the appropriate tasks for this detector.

In the following subsections we provide the results for coupling constraints in each of the main categories, namely decoupling from a package (section \ref{violations_package}), decoupling from a class (section \ref{violations_class}) and
decoupling from a method (section \ref{violations_method}).

\subsubsection{Detecting package decoupling violations} \label{violations_package}

In section \ref{DC_Package} we noted several cases where \DSpace packages should be decoupled from one another. Here we take one of those cases, build a detector for it and run the detector to determine if the coupling constraint is violated or not. We choose the requirement from figure \ref{fig:DSpaceArchitecture} that the Storage layer should only be accessed from the Business Logic layer.
%In this instance, the XML input to the detector is as follows:
%\begin{verbatim}
%<constraint
%  targetPackage="org.dspace.storage.rdbms"
%  allowedPackageList="org.dspace.administer; ... org.dspace.workflow;
%  message="only org.dspace can access org.dspace.storage" />
%\end{verbatim}
% Full list of allowed packages is: org.dspace.administer; org.dspace.authorize; org.dspace.browse; org.dspace.checker;
%org.dspace.content.dao; org.dspace.content; org.dspace.core; org.dspace.eperson; org.dspace.handle; org.dspace.search;
%org.dspace.storage.bitstore; org.dspace.workflow;
%The {\tt targetPackage} is the package that represents the Storage layer. The {\tt allowedPackageList} contains the packages in the Business Logic layer (there are 12 of them in total), with the implication that other packages should be decoupled from the {\tt targetPackage}.

When this detector was executed on \DSpacens, five distinct violations were found in four separate packages ({\tt app.statistics}, {\tt app.oai}, {\tt app.util} and {\tt app.webui.jsptag}). It is remarkable to find the essential architecture of the application being violated at all. Each of these violations represents an instance of the Application layer bypassing the Business Logic layer and accessing the Storage layer directly. In each case, the offending access was to Storage layer functionality required by the Application layer, but that was not exposed by the Business Logic layer.
%In three cases, the offending access was to the {\tt BitStreamManager} class.

These violations would be of great concern to a software architect, as they are signs that the architecture is starting to decay. Indeed, the two violations from the {\tt app.webui.jsptag} package also involved the duplication of an entire method in the Application layer, which is another clear indication of architectural decay.

Fixing these problems at this early stage is probably not a major challenge. The access to the desired functionality in the Storage layer should be exposed to the Application layer by the Business Logic layer, in keeping with the layering principle.

%\mel{We have a lot more detail about these violations, but I don't think it adds to the discussion to put them in.}
%\mel{We also have results for the METSExport CD, which was violated in seven places. Consider adding this as well.}

\subsubsection{Detecting class decoupling violations} \label{violations_class}

In section \ref{DC_Class} we saw the need to decouple the  {\tt BitstreamStorageManager} class from all \DSpace classes other than the {\tt Bitstream} class.
%The XML input to the \FindBugs detector in this instance is:
%\begin{verbatim}
%<constraint
%targetClass="org.BitstreamStorageManager"
%allowedClassList="org.Bitstream"
%message="Only org.Bitstream may access org.BitstreamStorageManager"\>
%\end{verbatim}

On creating and running the detector for this decoupling constraint, five violations were found. They originated in five separate classes, namely {\tt BitstreamDAO}, {\tt BrowseListTag}, {\tt Bitstream}, {\tt Cleanup} and {\tt ItemListTag}. In four cases the violation would appear to have been accidental, i.e., the programmer simply neglected to read the comment or failed to realise the import of the comment.

In the case of the violation in the {\tt BrowseListTag} class it is evident that the programmer wished to circumvent explicitly the authorisation required by the {\tt Bitstream} class, and so accessed the {\tt BitstreamStorageManager} class directly. This suggests that the design decision expressed in the original comment is too constraining for the programmers to work with. The reporting of a violated coupling constraint in this context suggests that the access to the {\tt Bitstream} and {\tt BitstreamStorageManager} classes may need to be redesigned.

\subsubsection{Detecting method decoupling violations} \label{violations_method}

In section \ref{DC_Method} the {\tt DAVEPersonEPerson} was described. The sole purpose of this class is to provide read-only access to an instance of the {\tt EPerson} class. We built a detector for this coupling constraint and executed it. No violations were discovered. To ensure that the detector was correct, we injected several random violations all of which were detected correctly.

%\mel{There are some problems here. The experimental results we have are for the Eperson example. However, in this case the DC is expressed not only as a comment but also by creating a new class to provide read-only access to the EPerson class. Hence no violations are found, which is hardly surprising. Three violations were injected, and these were all detected. This will make the discussion and presentation of this subsection different from the others, but maybe that's ok. Another possibility is to look at the Bundle.createBitstream example, but this would involve doing another experiment.}

\subsection{Discussion}\label{Discussion}

Our analysis of developer comments in \DSpace reveals a need for decoupling constraints. We found several cases where the developer wanted to constrain the future evolution of the program so as to avoid certain undesirable couplings, and expressed this as a comment. We only lay claim to the existence of this need; we have not tried to quantify it. We anticipate that our approach has a very high false negative rate. Most coupling constraints are probably not documented, and of the few that are, our blunt keyword search no doubt detected only a percentage of them.

We selected three coupling constraints to analyse further. A detector was developed that could detect violations of each of the chosen coupling constraints. We expected that in a well-regarded application like \DSpacens, no violations would be found. We were surprised to discover that two of the three coupling constraints were violated, and a total of ten violations were found. This is clear evidence that expressing coupling constraints in comments alone is not sufficient that that further tool support is necessary to ensure that coupling constraints are maintained during program evolution.

%This is partly due to the fact that, according to one case study \citep{Kozlov08}, open source software has a relatively low level of comments.

% ----- RELATED WORK -----
\section{Related Work}\label{RW}

In spite of its maturity, coupling remains a topic that attracts the interest of researchers. In this section we review related work in this field and demonstrate that coupling constraints, their detection, and their consequences, have not been addressed in the literature.

One of the earliest works in automated detection of object-oriented design problems is that of  Ciupke \citep{Ciupke99}. It aims to check a program for violations of object-oriented design principles, for example, to test if all fields are private in their class. These design principles are formulated as Prolog clauses and Ciupke shows how they can be detected in real applications. We also model the program being examined as a set of Prolog clauses and use Prolog queries to detect design violations. However, only one of the constraints Ciupke deals with is a coupling constraint, namely that a class should not know about its subclasses. He does not consider application-specific constraints which are the focus of our work.

Gu\'{e}h\'{e}neuc and Albin-Amiot \citep{Gueheneuc01} also deal with the detection of design problems. They argue that intra-class design problems have been well-studied and focus their attempts instead on detecting and correcting inter-class design defects. We share their viewpoint that ``inter-class design defects are difficult to define independently of the application and its context." However, they hypothesize further that design patterns embody quality architecture and that transforming structures that closely resemble design patterns to the normal pattern structure will improve architectural quality. The recognition of the problem of over-engineering caused by ``pattern happy" developers \citep{Kerievsky04} renders the first hypothesis suspect. Regarding the second hypothesis, patterns have many variations in their implementation structure, so a structure that is close to the prototypical pattern implementation may be perfectly valid in its context and not an appropriate for target for restructuring. By way of comparison, our approach is relatively agnostic in terms of design quality model, only assuming that in certain application-specific contexts, it is useful to decouple one program element from another.

The extent to which modules with poor structural measures (size, coupling, cohesion, inheritance) contribute to maintenance problems has been a topic of research for some time. Briand \emph{et al} performed an empirical evaluation of object-oriented design measures to determine their ability to predict fault-proneness \citep{Briand00}. They found many coupling and inheritance measures to be correlated with the probability of fault detection in a class. In later work, Koru and Tian analysed data from two large open-source projects and found that although there is indeed a correlation between modules with poor structural measures and change-proneness, the most change-prone modules were not those with the worst structural measures \citep{Koru05}. Yu \emph{et al} analyse intermodule coupling and show how the use of global variables in the Linux kernel has led to tighter coupling than was heretofore understood to be the case \citep{Yu04}\citep{Yu06}. They suggest that this coupling raises concerns about the long-term maintainablity of Linux. From our perspective, these various studies serve to confirm the importance of coupling.

Arisholm \emph{et al}. investigated the use of dynamic analysis to improve the measurement of intermodule coupling \citep{Arisholm04}. Static object-oriented coupling measurements do not take  polymorphism into account, and thus are prone to estimating incorrectly the true extent of interclass coupling. They demonstrate that dynamic measurements are better indicators of complexity than static measurements. In later work, Liu, Liu and Ana demonstrated that cheap, static analysis such as Rapid Type Analysis can compute dynamic coupling measures with almost perfect precision \citep{Liu06}. Our focus is on compile-time dependencies in order to reduce the ripple effect when one module is changed, so the use of static measures is more appropriate.

The concept of \emph{change coupling} is introduced by Ratzinger, Fischer, and Gall \citep{Ratzinger05}. Modules are changed coupled if they tend to be updated at the same time, according to  source code repository (e.g., CVS) data. Modules can be change coupled and have no detectable dependencies in the source code -- indeed this is by far the most insidious type of change coupling as it is undetectable by source code analysis.  More recent work by Eaddy {\emph at al} \citep{Eaddy08} demonstrates that non-modular crosscutting concerns tend to increase the number of defects in a program. This is is likely to be related to change coupling, in that modules that take part in a non-modular cross cutting concern can be expected to be changed coupled as well. Approaches based on source code analysis, such as ours, cannot detect this type of coupling. It can only be detected by an analysis of source code repository data.

Zaidman and Demeyer use coupling measures in combination with data mining techniques to detect key classes in \citep{Zaidman08}. They found that classes that are strongly coupled with others are likely to be key in terms of comprehending the software system. In this context, it should be noted that strong coupling is not necessarily bad. As explained by Martin \citep{Martin02}, a module such as an abstract class can have a high number of dependencies on it, but this is not a problem as long as the module is \emph{stable}, i.e., not subject to change. However, if an unstable module is similarly highly-coupled, it is likely to cause a strong ripple effect as each time the unstable module is changed, its dependant modules are likely also require updating. In our work we make no assumption that strong coupling is bad of itself, but rather enable the programmer/architect to define that certain application-specific couplings are to be avoided.

The recent work of Sarkar \emph{et al} \citep{Sarkar08} is relevant to ours in a number of ways. They point out that traditional metrics focus on the class as the module, but in large software systems it is the coupling across larger packages that is more important. The main contribution of their work is to propose and validate a set of metrics that characterizes large object-oriented software systems with regard to such dependencies. For example, they introduce a metric called the ``Module Interaction Index" that measures the extent to which modules are coupled only using their correct, published interface. An imperfect value for this measure indicates that undesirable inter-module coupling is taking place.  Another metric, the ``Not Programming to Interfaces Index,"  measures the extent to which client code uses subclasses directly, rather than through the interface provided at the root of the inheritance hierarchy. Preventing design decay in terms of these metrics is possible using coupling constraints.

%\mel{Add citation: ``the Ripple Effect \citep{Yau78}" {Rov08} needs to downloaded and read as well}

There is a large body of work in the field of Impact Analysis \citep{Rov08} which appears on the surface to be similar to our work. Impact analysis aims to discover the parts of a program that may be affected when a modification is performed. The analysis used may be static or dynamic, but in either case the goal is to find other modules whose behaviour might be affected by the modification. Our focus is rather on static, compile-time dependencies, which have no impact on behaviour. For example, the static dependency of a class {\tt A} on a class {\tt B} can be removed by creating an interface to {\tt B} and updating {\tt A} to depend on this new interface. This refactoring will not however affect the possibility of a change to the class {\tt B} having an impact on {\tt A}, as the runtime object structures are identical in both cases.
%\mikal{Impact on the behaviour I suppose? The point is precisely that some changes to B will not statically impact A thanks to the indirection so we do not want to confuse the reader here so it should be made clear what impact we are talking about here.}

There is some support for coupling constraints available in current software tools. In the Eclipse IDE \citep{Eclipse08}, it is possible to allow only limited access to classes/packages that are included from other projects. If the client code creates a dependency on a type or class that is not permitted, the Java compiler will report a warning or error. This is in effect a limited form of coupling constraint in that it can only be applied between a project and packages/classes that are from another project. For example, to limit an Eclipse project from accessing JRE classes outside of {\tt java.io.*} the following access rules should be added to the JRE classpath in the project:
\begin{verbatim}
Accessible : java/io/*
Forbidden : **
\end{verbatim}
Another example is the import control feature provided with {\tt CheckStyle} \citep{CheckStyle08}, an open source tool that checks Java code for a variety of coding problems. The import control feature checks that all import statements follow the layering and import rules defined in a project XML file. The motivation behind this tool is similar to ours: to prevent a programmer carelessly creating an undesirable dependency on a class in a package. Our work goes much further than this, by considering decoupling between all program elements, not only packages.

Finally, there is of course some support for coupling constraints in the programming languages through various mechanisms to restrict the visibility of program elements \citep{pattern:ardourel1}. This support is unfortunately not sufficient as will be shown in section \ref{Eval} through the comments that programmers felt were needed to warn about unwanted couplings.

%\mel{Cite also ``Can we avoid high coupling?'' from ECOOP 2011 -- claims that high coupling cannot be avoided!}

Following  \citep{pattern:ardourel1}, we have called our graphs``access graphs" because our fundamental relation, \emph{uses}, binds program elements to the scopes which use (access) them. Our access graphs are simpler, though, than those of \citep{pattern:ardourel1} because the \emph{uses} relation abstracts various kinds of accesses. All that matters to us here, is that the name of a program entity appears or not in some scope, thereby exposing or not the scope to changes of the program entity.

%------------------------------------------------------------------------------------
\section{Conclusions and Future Work}\label{CnFW}
%------------------------------------------------------------------------------------
In this paper we have:
\begin{itemize}
\item defined the concept of static dependency as the occurrence of a name in a scope,
\item defined the concept of access graph to reason about static dependencies,
\item defined a logical framework to express coupling constraints that forbid some static dependencies,
\item demonstrated the need for coupling constrains by finding occurrences of them in comments in Dspace,
\item expressed these constraints using our language,
\item found several violations of these constraints in DSpace using \FindBugs and our own tool, Lutin.
\end{itemize}
We thus draw the conclusion that coupling constraints should be made explicit so that they are both easy to understand by human developers and supported by tools that can detect their violations.

This should greatly help software designers analyze the impact of changes as advised for instance by \citep{Martin02}. They will try and keep them local by hiding the scopes which are expected to change from scopes which change at a different pace. The hidden scopes may still be used indirectly from facades or trough abstractions of their types using dynamic binding (which creates no static dependency). The point of using explicit coupling constraints is that they point out precisely where indirections and abstractions are needed to avoid over-engineering. Finally, explicit coupling constraints help prevent the decay of software architectures by pointing out where coupling constraints are not enforced any more.

Future work includes the semi-automatic control of refactoring transformations to enforce coupling constraints and application to design patterns.

%Future work includes transformation and elaboration of the link with Design Patterns.

% The potential of tool support is considerable.

%Is automation of the process of adding coupling constraints possible?

%Criticisms of this work:
%\begin{itemize}
%
%\item coupling constraints must be stated explicitly by the programmer or system architect. The benefits may outweigh the cost, but approaches that involve extra annotation are invariably slower to gain acceptance. We plan to investigate the automated derivation of coupling constraints from other sources such as CVS repositories and program documentation.
%
%\item coupling constraints can only be used to express decoupling from elements/interfaces defined in the programming language itself. In one of the examples we found, the decoupling expressed was from parts of a database interface, accessed through RMI-packaged (terminology?) requests. Static checking of this kind of decoupling is impossible, as the types involved are unknown. (Example is in deleted text at the end of the paper)
%
%\item ...
%
%\end{itemize}

\bibliographystyle{plainnat}
\bibliography{paper}

\end{document}